\documentclass{article}
\usepackage[T1]{fontenc} % add special characters (e.g., umlaute)
\usepackage[utf8]{inputenc} % set utf-8 as default input encoding
\usepackage{ismir,amsmath,cite,url}
\usepackage{graphicx}
\usepackage{color}
\usepackage{glossaries}

\usepackage{subcaption}
\usepackage{dsfont}

\usepackage{lineno}

\DeclareMathOperator{\sign}{sign}
\DeclareMathOperator{\clip}{clip}

\newacronym{mir}{MIR}{Music Information Retrieval}
\newacronym{cw}{C\&W}{Carlini \& Wagner}
\newacronym{lime}{LIME}{Local Interpretable Model-agnostic Explanations}
\newacronym{cnn}{CNN}{Convolutional Neural Network}

\title{On the Veracity of Local, Model-agnostic Explanations in Audio Classification: \\
Targeted Investigations with Adversarial Examples}

\multauthor
{Verena Praher$^{1*}$ \hspace{1cm} Katharina Prinz$^{1*}$\thanks{* Equal contribution.} \hspace{1cm} Arthur Flexer$^1$ \hspace{1cm} Gerhard Widmer$^{1,2}$} {
  $^1$ Johannes Kepler University Linz, Austria \\
$^2$ LIT AI Lab, Linz Institute of Technology, Austria \\
{\tt\{verena.praher,katharina.prinz\}@jku.at}}

\def\authorname{V. Praher, K. Prinz, A. Flexer, and G. Widmer}

\begin{document}
\maketitle

\begin{abstract}

Local explanation methods such as LIME have become popular in MIR as
tools for generating post-hoc, model-agnostic explanations of a
model's classification decisions. The basic idea is to identify
a small set of human-understandable features of the classified example that are most influential on the classifier's prediction.
These are then presented as an explanation.
Evaluation of such explanations in publications often resorts to
accepting what matches the expectation of a human without actually being able to verify if what the explanation shows is what really caused the model's prediction.
This paper reports on targeted investigations where we try to get more insight into the actual veracity of LIME's explanations in an audio classification task.
We deliberately design adversarial examples for the classifier, in a way that
gives us knowledge about which parts of the input are potentially responsible
for the model's (wrong) prediction. Asking LIME to explain the predictions for these adversaries permits us to study whether local explanations do indeed detect these regions
of interest. We also look at whether LIME is more successful in finding perturbations that are more prominent and easily noticeable for a human. 
Our results suggest that LIME does not necessarily manage to identify the most  relevant input features and hence it remains unclear whether explanations are useful or even misleading.

\end{abstract}

%----------------------------------------------------------------------

\section{Introduction}\label{sec:introduction}

%% Problem Statement

With the rise of deep learning methods used in \gls*{mir}, also the desire for explaining the decisions made by such models has increased. A plethora of explanation methods (``explainers'') have been originally developed for text or image data and adapted to the audio domain~\cite{mishra2018a,mishra2018eusipco}, or specifically introduced for \gls*{mir} systems~\cite{mishra2019GANs}. Most notably, different versions of \gls*{lime}, a post-hoc explainer~\cite{ribeiro2016lime}, 
have been used to explain models in a variety of \gls*{mir} tasks~\cite{mishra2017local, chettri18spoofing, Haunschmid2019TwoLevel, Haunschmid2020arxiv, Haunschmid2020MML, melchiorre2021ecir, mishra2020reliable}. 

Evaluation of explanations is known to be a hard task, due to a lack of agreement on what constitutes a ``good explanation''~\cite{lipton18mythos} and, consequently, what may serve as ground truth for measuring the quality of an explanation. Explanations are often evaluated visually and the applied ``metric'' is whether the highlighted input parts
% parts highlighted by the explanation 
appear relevant to the human observer (``confirmation bias'')~\cite{adebayo2018sanitychecks}.

We propose evaluating local explanations by exploiting a known weakness of deep neural networks: adversarial examples. By adversarially perturbing examples in a way that a system's original prediction is changed, we know exactly what in the input caused the erroneous (\textit{adversarial}) prediction and can use this as the ``ground truth'' that the explanation method should recover.

%% Related Work

We are not the first to investigate the relationship between adversarial examples and interpretability. Slack et al.~\cite{slack20fooling} show that a racially biased model can be attacked in a way that the (biased) prediction remains unchanged but the explanation appears as if the model did not base the prediction on sensitive attributes. This work is related in the sense that it highlights weaknesses of post-hoc explanation methods (including \gls*{lime}).

Closest to our work are Göpfert et al.~\cite{goepfert2019recovering}, who use ``localised'' adversarial attacks that target only selected segments of an image, and investigate the ability of different explainers (including \gls*{lime}) to recover the affected part. In their experiments \gls*{lime} outperforms the other explainers, which supports our choice of \gls*{lime} as an explainer that deserves a more careful and critical investigation.

%% Contributions

We perform four experiments with purposefully designed adversarial examples in order to obtain more insight about strengths and weaknesses of LIME-based explanations in an audio classification task. As a test bed, %test vehicle
we have chosen a DNN-based singing voice detection model~\cite{schlueter2018zeromean} that currently defines the state of the art on that task, and
is suited to study the quality of explanations for several reasons: it addresses
a clearly defined task (as opposed to genre classification for example), it has been used for demonstrating explanations before~ \cite{mishra2017local, mishra2020reliable}, and it has the interesting previously documented property that it can be confused by directly drawing on the spectrogram~\cite{schlueter2017_phd}.

%----------------------------------------------------------------------

\section{Local Interpretable Model-agnostic Explanations}
\label{sec:lime}

LIME is an algorithm for explaining individual predictions (\textit{local}) for any black box machine learning model (it is \textit{model-agnostic}). The general idea of the algorithm is to train a simpler model $g$ that approximates the neighborhood of the prediction $f(x)$ that we want to explain. The first step is to derive an interpretable representation depending on the input domain, e.g. the presence or absence of super pixels\footnote{(perceptually) grouped pixels} in the image domain.

Let $x \in \mathds{R}^{d}$ (in the case of spectrograms, $d=T\times F$) be the representation in the original input domain and $x' \in \{0, 1\}^{d'}$ be the interpretable representation where 0 and 1 denote the absence and presence of an interpretable feature, respectively. In the next step we sample $N_s$ instances around $x'$ by randomly generating vectors of length $d'$ containing 0's and 1's. Each generated instance $z'$ is mapped back to the input domain and fed through the original model $f$. $z \in \mathds{R}^d$ looks like the original input example with all parts that are 0 in $z'$ occluded (e.g. gray values in the image domain). The $N_s$ instances $z'$ and the corresponding predictions $f(z)$ constitute the input to the explanation model $g$~\cite{ribeiro2016lime}. The most commonly used explainers are linear models of the form\footnote{In the original paper~\cite{ribeiro2016lime}, the explainer had the form $g(z') = w_g z'$, mistakenly
omitting the 
intercept $b$. Inspecting 
the source code shows that the intercept was trained  as well:
\url{https://github.com/marcotcr/lime/}}

\begin{equation}\label{eq:linear_model}
g(z') = b + w_g z'.
\end{equation}

When training the linear explanation model, the generated instances are weighted via an exponential kernel decaying with increasing distance to the input example. This distance (and based on it the weight) is computed between the binary representation of the examples being explained (all $1$'s) and each of the instances in the neighborhood\footnote{The original paper~\cite{ribeiro2016lime} stated that the distance for weighting the examples was computed between the images directly, but the source code shows that it is actually computed between the binary representations, which was later confirmed by the author.
}.

A requirement for an explanation is that it be \textit{locally faithful}: we expect the explainer $g$ to approximate $f$ in the neighborhood of $x$. The faithfulness is measured by the fidelity score, which is computed by the coefficient of determination ($R^2$) of the linear model's predictions~\cite{ribeiro2016lime}, and ranges from 0 to a perfect score of 1, with the exact interpretation of the range of values being somewhat unclear.

Different flavours of LIME have been used for explaining predictions in a variety of audio classification and tagging tasks. The general LIME algorithm was kept the same and the difference is the type of interpretable features. Mishra et al.~\cite{mishra2017local} proposed the first adaption of LIME for MIR, termed SoundLIME (SLIME), which segments the spectrogram into time, frequency, or time-frequency segments. SLIME was demonstrated on the task of singing voice detection~\cite{schlueter2015svd} and used for analysing a replay spoofing detection system~\cite{chettri18spoofing}. Haunschmid et al. used other types of interpretable features (super pixels~\cite{Haunschmid2019TwoLevel}, source separation estimates~\cite{Haunschmid2020arxiv, Haunschmid2020MML}) for explaining the predictions of a variety of models, including music taggers~\cite{Haunschmid2020arxiv, Haunschmid2020MML} and a content-based music recommender system~\cite{melchiorre2021ecir}.
Mishra et al.~\cite{mishra2020reliable} proposed different content types for replacing the ``grayed out'' segments (e.g. \textit{zero}, the \textit{min} and the \textit{mean} value of the spectrogram) and found the mean value to work best in their setting.
For our experiments we will be using time-frequency segments~\cite{mishra2017local, mishra2020reliable}.
The hyperparameters for the LIME algorithm are described in Section~\ref{sec:lime_params}.

%----------------------------------------------------------------------

\section{Adversarial Attacks}
\label{sec:Adversarial Attacks}

In this section, we briefly describe a way to compute adversarial examples for a singing voice detection system. The goal is to obtain perturbations which do not affect how a human perceives a particular example, yet simultaneously change the prediction of the system, i.e., transform ``Singing Voice'' to ``No Singing Voice'' or vice versa. To realise this, we use an adversarial attack that was originally proposed for audio~\cite{Carlini2018STT}, and was previously applied in \gls*{mir} to attack an instrument classification system and a music recommender~\cite{Prinz2020Arxiv}. The attack, subsequently denoted by \gls*{cw}, assumes a white-box scenario, i.e.\ full knowledge about a model and its parameters.

As \gls*{cw} is an iterative targeted attack, we use it to decrease the loss of a system with respect to a new (wrong) target prediction, in multiple iterations. Following the notation in~\cite{Prinz2020Arxiv}, let $f$ be a system and $x \in \mathds{R}^d$ its input, i.e.,\ a specific prediction is denoted by $f(x)$. Also, let $\delta_{ep} \in \mathds{R}^d$ be an adversarial perturbation at %a certain 
iteration $ep$, and an adversarial example $x + \delta_{ep}$. Furthermore, we denote the system loss by $L_\mathrm{sys}$, and the target prediction by $t$. To restrict the adversarial perturbation, \gls*{cw} minimises a weighted sum of the squared L2-norm of $\delta$ and a system loss, resulting in an optimisation objective as follows~\cite{Carlini2018STT,Prinz2020Arxiv}:

\begin{align}
\label{eq:cw}
    L_\mathrm{total} &= \|\delta_{ep}\|^2_2 + \alpha * L_\mathrm{sys}(f(x + \delta_{ep}), t), \nonumber \\
    \delta_{ep+1} &= \clip_\epsilon (\delta_{ep} - \eta * \sign(\nabla_{\delta_{ep}} \ L_\mathrm{total})).
\end{align}

Note that $\nabla_{\delta_{ep}}$ is the gradient w.r.t. to ${\delta_{ep}}$, and updates are performed based on the $\sign$ of the gradient and the multiplicative factor $\eta$. By applying $\clip_\epsilon$, an adversarial perturbation is always in the range of $[-\epsilon, \epsilon]$, and $\alpha$ is used to balance the magnitude of the perturbation and the system loss w.r.t. the target prediction.

In the following experiments, we will compute adversarial examples both for raw audio waveforms and for input spectrograms. In either case, we will always analyse the resulting perturbations in the time-frequency domain. For adversaries computed for the waveform we compute the time-frequency representation and subtract the original example to obtain the adversarial perturbation in the time-frequency domain. For adversarial examples computed for the spectrogram directly, time-frequency information is already available.

%----------------------------------------------------------------------

\section{Experimental Setup}
\label{sec:setup}

This section describes the experimental setup, including the data we use, details about the singing voice detection system and hyperparameters for both the adversarial attacks and computing the explanations.

\subsection{Data}
\label{subsec:data}
To train the singing voice detection system proposed by Schlüter and Lehner, we use a subset of the data used in their work~\cite{schlueter2018zeromean}, namely the openly available \textit{Jamendo} dataset\cite{Ramona2008Jamendo}. The dataset consists of 93 songs totalling around 6 hours of music, with a proposed training / validation / test split of 61 / 16 / 16 files respectively. Each audio file has a sampling rate of 44.1kHz.%, and is released with a Creative Commons licence. 

The annotations for the Jamendo dataset are provided with sub-second granularity~\cite{schlueter2015svd}, and indicate the presence / absence of singing voice. The proportion of singing voice (``sing'') in comparison to non-singing voice (``no sing'')  is close to $50:50$ in all three data splits.

\subsection{Singing Voice Detector}
\label{subsec:singing_voice_detector}
For subsequent experiments, we adapt the singing voice detection system introduced by Schlüter and Lehner~\cite{schlueter2018zeromean}. We use the proposed architecture of the \gls*{cnn} and deploy the training routine with unaltered hyperparameters on the Jamendo dataset. Prior to training, the data is resampled to 22.05kHz and then used to compute magnitude Mel spectrograms (frame length 1024, hop size 315 samples and 80 Mel bands)~\cite{schlueter2018zeromean}. The Mel spectrograms are logarithmically scaled and normalised per frequency band to have zero mean and unit variance over the training data. 

The \gls*{cnn} is trained on spectrogram excerpts with a length of 115 frames, for which the target prediction corresponds to the ground-truth annotation for the central frame~\cite{schlueter2018zeromean}. During training, we use a mini-batch size of 32 and Adam to optimise the Binary Cross Entropy loss for 40.000 update steps. We start with a learning rate of 0.001 and scale it by 0.85 every epoch. To support generalisation, dropout and %the proposed 
data augmentation~\cite{schlueter2015svd} is used;
for a more detailed description of the training procedure, hyperparameters and the CNN architecture, %architecture of the CNN, 
we refer to~\cite{schlueter2018zeromean}.

The final classification error of the singing voice detector (in percent) on the Jamendo test data, given as the mean $\pm$ standard deviation over 5 different runs, is $11.54 \pm 0.96$. Furthermore, recall and specificity over 5 runs are $89.61 \pm 1.71$ and $87.46 \pm 1.00$ respectively. The metrics are computed based on binary predictions of the network, which are obtained after applying a median filter and a threshold tuned on validation data (cf. \cite{schlueter2018zeromean}). Note that exact reproduction of previously reported errors, namely 8.0 in \cite{schlueter2015svd} (with a slightly different architecture) and 5.5 in \cite{schlueter2018zeromean},
is difficult as Schlüter and Lehner train on a larger (in-house) dataset, whereas we only use publicly available data. Additionally, non-determinism (e.g., via data augmentation or initial weights) can influence the performance of a model noticeably.

In the following experiments, let the singing voice detection system be $f$, and $f(x)$ the numeric model output for class ``sing''. The final classification is made by checking whether $f(x)$ is above (``sing'') or below (``no sing'') a threshold that is optimised on the validation set, and is equal to 0.51 for our system. 

\subsection{Hyperparameters for the Adversarial Attack}
\label{subsec:hyperparametes_adv_attack}
Equation \eqref{eq:cw} shows the hyperparameters we need to determine before attacking the singing voice detector. 
The maximum number of iterations in which we try to find adversarial perturbations that change the prediction of an excerpt is set to 1000 in our experiments. Other hyperparameters -- clipping factor $\epsilon$, update factor $\eta$ and weight factor $\alpha$ -- 
are tuned on the validation set of Jamendo, and chosen as the setting that results in the highest number of successful adversarial perturbations, i.e., changed the most predictions out of all audio excerpts in the validation data. 
For the attack on raw audio we use $\epsilon = 0.01$, $\eta = 0.0003$ and $\alpha = 2$; for the attack on the spectrograms we take $\epsilon = 0.1$, $\eta = 0.0005$ and $\alpha = 15$. Due to the binary nature of the singing voice detection task, the target $t$ for each excerpt is chosen to be the opposite of its original prediction. Note that we find successful perturbations for 28.4\% of all excerpts for attacks on the spectrogram, and for 60.5\% of excerpts for raw audio.

\subsection{Hyperparameters for the Explanations}
\label{sec:lime_params}

As mentioned above the LIME algorithm has a set of hyperparameters that have to be chosen. We segment the spectrogram ($80$ frequency bins ($F$) $\times$ $115$ time frames ($T$)) into 5 time and 4 frequency segments,
respectively, resulting in 20 ``human-interpretable'' features of size $20$ frequency bins $\times$ $23$ time frames.
Repeating the calculation of an explanation might lead to different results due to the randomness in the neighborhood generation when the number of generated instances is too small. A preliminary experiment as described in~\cite{mishra2017local,mishra2020reliable} on a subset of the explanations, suggests that $2^{13} = 8192$ instances are sufficient to generate stable explanations, which is why we set $N_s=8192$ for all our experiments. We also investigated different ways of replacing
explanation segments~\cite{mishra2020reliable} (\textit{zero}, \textit{min}, and \textit{mean}), but as the results appeared
similar, we will focus on reporting the results for the \textit{mean} content type. For weighting the instances to train the explainer we use the cosine distance function and an exponential kernel with a kernel width of $0.25$.

%----------------------------------------------------------------------

\section{Evaluation of Explanations}%{Explanation Evaluation using Adversaries}
\label{sec:evaluation}

To investigate to what extent local explanation methods can find the underlying reason for a particular classification by a system, we perform four experiments. In the first, we demonstrate how LIME can help detecting rather obvious causes for a prediction that are due to manually altered spectrograms. The goal of the remaining experiments is to quantitatively evaluate the ability of LIME to detect adversarial perturbations: in the second experiment we try to evaluate whether the segments that \gls*{lime} highlights actually caused a prediction; in the third, we analyse if the explanations can detect the correct location of the adversarial perturbation if it only affects few segments of the input spectrogram; and in the fourth, we compare the fidelity scores for correct and incorrect explanations. 

Due to the computation time needed for obtaining explanations, we subsequently limit the number of analysed excerpts for each song in the test set. For the first experiment, we manually alter a single random excerpt per song with an original classification of ``no sing''. In the setting of the second, third and fourth experiment, for each of the 16 test songs, we randomly select 10 adversarial excerpts which were originally classified as ``sing'' and another 10 excerpts originally classified as ``no sing''. 
We conduct all experiments with adversarial attacks computed both on raw audio and on spectrograms, as we want to account for the fact that attacks on raw audio might use implicit constraints that are difficult to detect for LIME (which operates on spectrograms).

To allow reproducibility, we publish code\footnote{\url{https://github.com/CPJKU/veracity}} for all subsequent experiments.

\subsection{Explaining Manually Altered Spectrograms}
\label{sec:horse}

\begin{figure}[ht]
    \begin{subfigure}{1\columnwidth}
      \centerline{
 \includegraphics[width=\columnwidth]{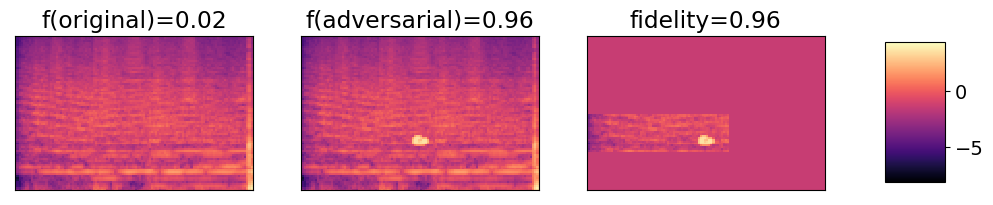}}
    \end{subfigure}
    \begin{subfigure}{1\columnwidth}
      \centerline{
 \includegraphics[width=1\columnwidth]{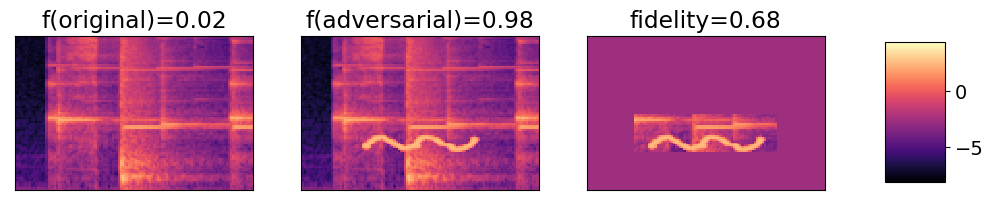}}
    \end{subfigure}
    \begin{subfigure}{1\columnwidth}
      \centerline{
 \includegraphics[width=1\columnwidth]{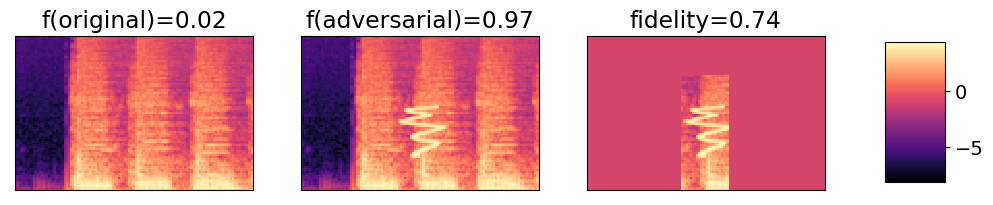}}
    \end{subfigure}
 \caption{Different examples of spectrograms for which the model's original prediction is ``no sing'' (left column) but drawing certain symbols (middle) leads to a change to ``sing'' with a model output $f$. The rightmost column shows the top 3 
 interpretable features (which may be directly adjacent and thus look like one segment) picked by the LIME algorithm, and their fidelity.
 }
 \label{fig:horses}
\end{figure}

In our first experiment, we exploit the fact that the predecessor of the singing voice detector we use (cf. \cite{schlueter2015svd}) could easily be fooled by drawing on the input spectrogram\footnote{\url{https://github.com/f0k/singing_horse}}~\cite{schlueter2017_phd}.
After choosing a set of excerpts originally classified as ``no sing'', we adapt drawings on the respective spectrograms until the prediction is changed to ``sing''. Then we ask LIME to explain these new predictions. Figure~\ref{fig:horses} shows that the explainer can correctly identify the manually altered parts of the spectrogram as the cause for the wrong prediction. Examples like these, supported by relatively high fidelity scores, might make us think that LIME knows what is going on and make us trust these explanations. We could stop our evaluation here and put trust in our model because the explanations highlight seemingly correct behaviour~\cite{mishra2017local}, look meaningful~\cite{Haunschmid2019TwoLevel}, or match the expected behavior~\cite{Haunschmid2020MML}, but we will continue with a more careful investigation.

\subsection{Explaining Predictions on Adversarial Examples} 
\label{sec:explaining_adversaries}

In this set of experiments, we first compute explanations for the predictions obtained for each of the selected adversarial excerpts. Every explanation consists of a list of interpretable features (time-frequency segments) and their corresponding weights. The weight of a particular feature should indicate the importance of the feature for making a certain prediction~\cite{ribeiro2016lime, mishra2017local}. To evaluate whether the explanation actually selects the features that are responsible for the (now wrong) prediction, we could naively check if the segments selected by LIME correspond to regions of the adversarial perturbation that have the highest magnitudes (defined via the L2 norm).

However, as we do not know which part of a perturbation really led to a misclassification, i.e., whether parts of a perturbation with the highest magnitude have the most influence or not, this could lead to false conclusions. 

To circumvent this, we follow another approach: we split the adversarial perturbation into the same time-frequency segments that LIME is using as interpretable features;
we then selectively add those $k$ segments of the perturbation to the input spectrogram that coincide with the features identified as an explanation. If the selected features are actually explaining the prediction, we expect the partial perturbation to be sufficient to achieve the same change of classification, or what we %subsequently
call \textit{label flip}, as the full perturbation did. However, using $k=3$ (which seems a reasonable number of segments to present to a user~\cite{mishra2017local}) flips only $21 - 53 \%$ of the predictions (see Table~\ref{tbl:preview}).

\begin{table}
\centering
    \begin{tabular}{l c c}
    ~               & Waveform & Spectrogram       \\ \hline
    ``no sing'' $\rightarrow$ ``sing'' & 53 \% & 41 \%         \\
    ``sing'' $\rightarrow$ ``no sing'' & 21 \% & 32 \%         \\
    \end{tabular}
    \caption{Percentage of label flips when using only $k=3$ segments of an adversarial perturbation, based on the most relevant features chosen by LIME. Columns denote whether perturbations were computed for the waveform / spectrogram, rows show original $\rightarrow$ target prediction.}
    \label{tbl:preview}
\end{table}

\begin{figure}
\captionsetup[subfigure]{justification=centering}
    \begin{subfigure}{.24\textwidth}
      \centerline{
 \includegraphics[width=1\columnwidth]{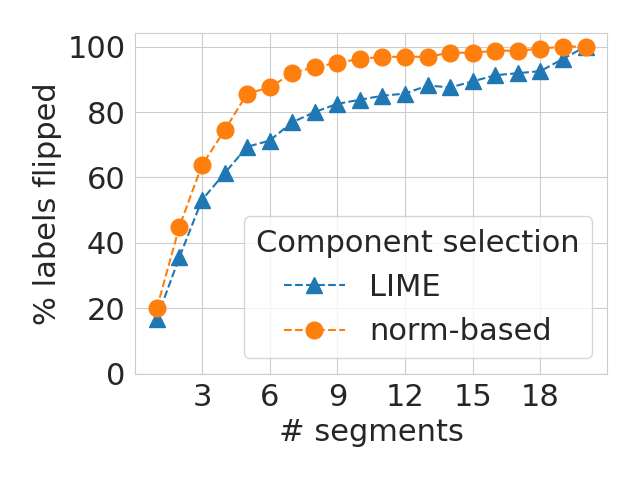}}
  \caption{``no sing'' $\rightarrow$ ``sing'' \\ (waveform)}
    \end{subfigure}%
    \begin{subfigure}{.24\textwidth}
      \centerline{
 \includegraphics[width=1\columnwidth]{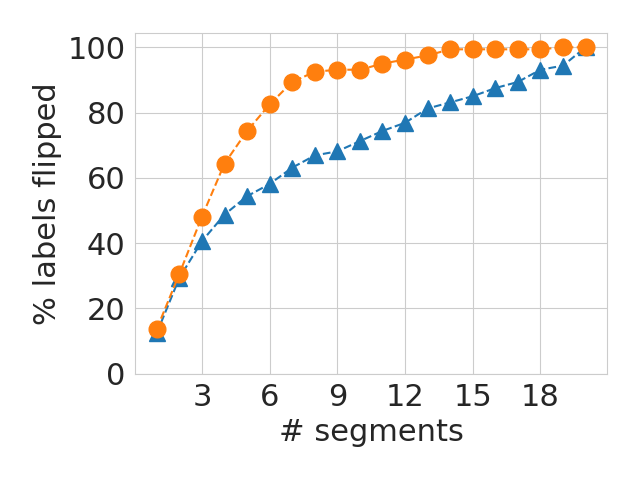}}
  \caption{``no sing'' $\rightarrow$ ``sing'' \\ (spectrogram)}
    \end{subfigure}
    \begin{subfigure}{.24\textwidth}
      \centerline{
 \includegraphics[width=1\columnwidth]{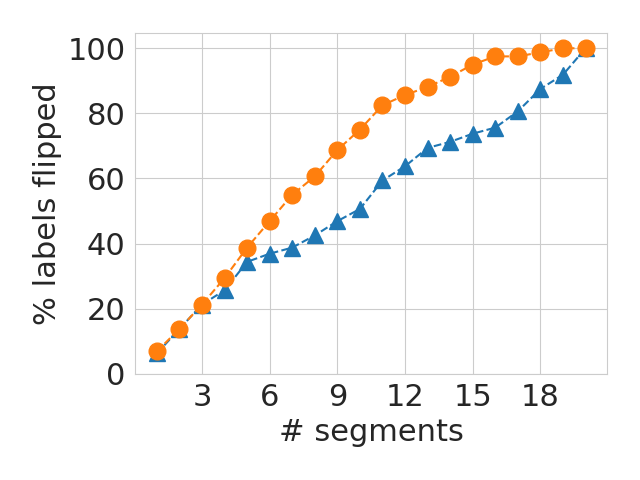}}
  \caption{``sing'' $\rightarrow$ ``no sing'' \\ (waveform)}
    \end{subfigure}%
    \begin{subfigure}{.24\textwidth}
      \centerline{
 \includegraphics[width=\columnwidth]{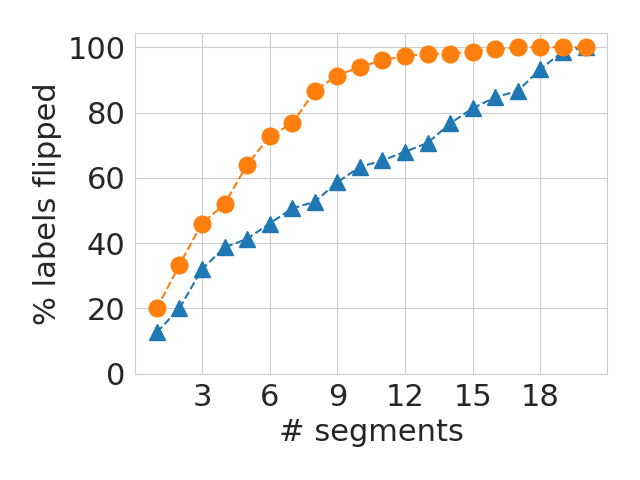}}
 \caption{``sing'' $\rightarrow$ ``no sing'' \\ (spectrogram)}
    \end{subfigure}%
 \caption{Percentage of classifications we can flip (y-axis) by adding only $k$ segments (x-axis) of an adversarial perturbation to an input. Choosing which $k$ segments are used is either based on LIME explanations (triangles), or on the norm of a perturbation (circles).
 }
 \label{fig:flipped}
\end{figure}

%% first sub experiment (1)
We therefore next look at whether larger numbers $k$ of segments are able to flip more of the predictions. We use either LIME for selection of segments or choose the $k$ perturbation segments with the highest magnitudes.
We do this for $k \in [1, ..., 20]$ and report the results in Figure~\ref{fig:flipped}.
For almost all $k$ and across different settings we are more often successful in flipping the label when segment selection is based on segment magnitude rather than on LIME results. For LIME-based segments it is necessary to choose almost all 20 available segments to achieve flip rates close to $100\%$. This result suggests that it is indeed possible to explain predictions with larger numbers of $k$ interpretable features, but only when using the norm of a perturbation (which in a real application we would not know) to select the features and not when using LIME for the selection.

%% second sub experiment (2)
Finally, we analyse whether taking all segments that receive a positive weight from LIME can flip a prediction, as positive weights should correlate positively with the explained class~\cite{mishra2017local}. For this analysis, we no longer add a fixed number of segments of a perturbation to each excerpt, but instead select all $k$ segments of a perturbation for which the corresponding interpretable feature received a positive weight. This number of segments $k$ can be different for each excerpt we look at. Additionally, we choose the same number of $k$ segments based on the largest magnitude, and compare how often this results in label flips. The results of this experiment are summarised in Table~\ref{tbl:positive}. Again, we observe that taking partial perturbations is more successful when selecting the segments with the highest magnitude, as opposed to the segments most relevant for LIME.

\begin{table}
    \begin{tabular}{l cc cc}
    ~               & \multicolumn{2}{c}{Waveform} & \multicolumn{2}{c}{Spectrogram}       \\
    ~               & LIME & norm & LIME & norm \\ \hline
    ``no sing'' $\rightarrow$ ``sing'' & 86 \%    & 99 \%  & 91  \%  & 98  \%         \\
    ``sing'' $\rightarrow$ ``no sing'' & 49  \%   & 74  \% & 44  \%   & 55  \%         \\
    \end{tabular}
    \caption{
    Percentage of label flips when adding all perturbation segments that contributed positively according to LIME, compared to using the same number of segments $k$ selected based on the norm.
    }
    \label{tbl:positive}
\end{table}

These results suggest that LIME does not pick the input features that are most relevant for making a particular prediction.

\subsection{Explaining ``Localised'' Perturbations} 
\label{sec:explaining_localized}

Considering the results in the previous section, one might argue that an adversarial perturbation can be present in the whole spectrogram, which could make
it hard for LIME to decide which segments are most relevant. We make use of the finding that it is often sufficient to add only a small number of selected segments of the perturbation to flip a label, and conduct an additional experiment where we analyse how often LIME detects those segments.

This is similar to~\cite{goepfert2019recovering}, where adversarial perturbations were constrained beforehand such that only selected regions of the input could be modified, but in contrast we do not restrict the perturbations directly. Instead we refine our adversarial perturbations by first splitting them into segments that correspond exactly to the time-frequency segments that LIME uses. We then use only $k$ such segments of a perturbation, where the segments themselves are chosen based on the highest magnitude.

We then identify the subset of adversarial excerpts whose label can be flipped by only adding $k \in {1, 3, 5}$ segments of the perturbation. 
These subsets are created for the four different settings, namely the two types of adversarial attacks (waveform / spectrogram), and for the two possible label flips (``sing'' to ``no sing'' and vice versa). This leads to $3 * 2 * 2 = 12$ subsets of adversarial excerpts.

\begin{figure*}
\begin{subfigure}{\columnwidth}
      \centerline{
 \includegraphics[width=\textwidth]{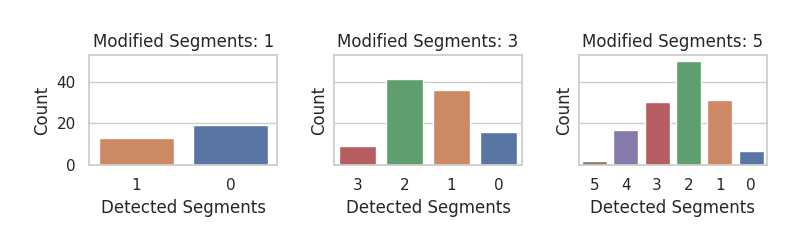}}
 \caption{no sing $\rightarrow$ sing (waveform)%(32, 102, 137 examples are shown, respectively)
 }
\end{subfigure}
\begin{subfigure}{\columnwidth}
      \centerline{
 \includegraphics[width=\textwidth]{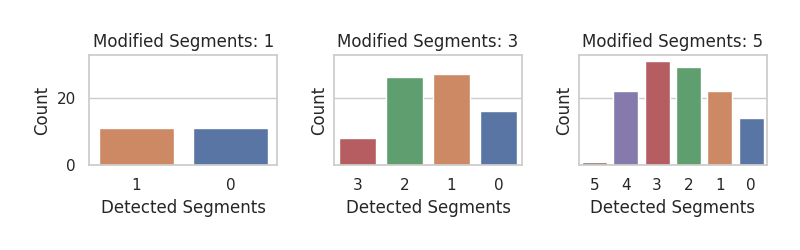}}
 \caption{no sing $\rightarrow$ sing (spectrogram) %(22, 77, 119 examples are shown, respectively)
 }
\end{subfigure}

\begin{subfigure}{\columnwidth}
      \centerline{
 \includegraphics[width=\textwidth]{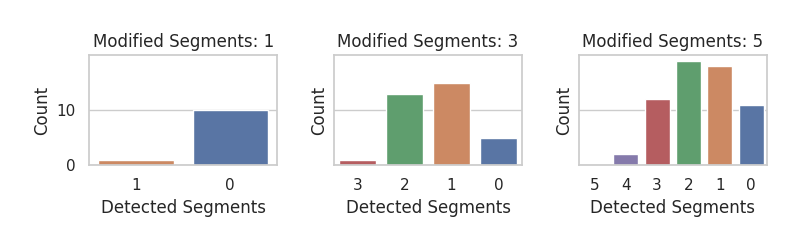}}
 \caption{sing $\rightarrow$ no sing (waveform) %(11, 34, 62 examples are shown, respectively)
 }
\end{subfigure}
\begin{subfigure}{\columnwidth}
      \centerline{
 \includegraphics[width=\textwidth]{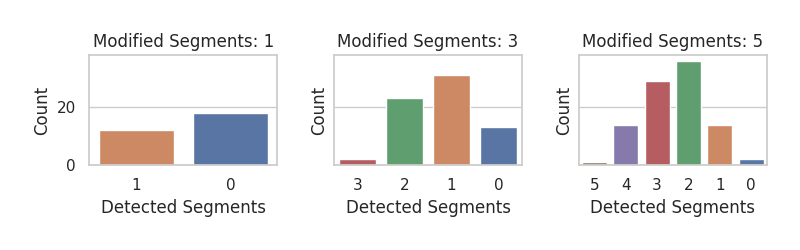}}
 \caption{sing $\rightarrow$ no sing (spectrogram) %(30, 69, 96 examples are shown, respectively)
 }
\end{subfigure}

\caption{The number of segments that are correctly identified when adding $k$ perturbed segments to the original input and asking LIME for an explanation containing the same $k$ number of interpretable features.}
 \label{fig:subset_perturbation}
\end{figure*}

After determining this subset of adversarial excerpts with partial (``localised'') perturbations, we once again compute explanations with LIME. Here we set the number of interpretable features that LIME should display to $k$, i.e. the number of segments we previously added as localised perturbations. Since the time-frequency segments of LIME and the partial perturbations align, we can then examine how often LIME correctly identifies the regions causing a particular prediction. Figure~\ref{fig:subset_perturbation} shows the result of this experiment, for all 12 adversarial subsets. Each plot depicts how many of the $k$ segments that were added as adversarial perturbation, and that hence were crucial for a particular classification, are correctly detected by LIME. It is easiest to interpret results with only one modified segment (left column), since we know exactly what the correct explanation should look like. Overall, for less than half of the excerpts is the correct segment presented as an explanation. For 3 and 5 modified segments we can also check how many segments are correctly identified, and we can observe that it is rarely all of them and for some settings none of them. This result suggests that even when the cause is quite localised and aligns with the segments that we are using as interpretable features, LIME is rarely able to recover all affected features.

\subsection{Are Fidelity Scores Reliable?}
\label{subsec:fidelity_reliable}

As Figure~\ref{fig:horses} in Section~\ref{sec:horse} suggested previously, ``correct'' explanations are often accompanied by relatively high fidelity values. We do not know, however, whether we can use fidelity to determine if an explanation highlights the segments of the input that are really the most relevant for a prediction, and it has been demonstrated that a high-fidelity explanation of a black box model might not reflect the actual cause for a prediction~\cite{lakkaraju20misleading}.

\begin{figure}
\captionsetup[subfigure]{justification=centering}
\begin{subfigure}{0.5\columnwidth}
 \centerline{
 \includegraphics[width=\textwidth]{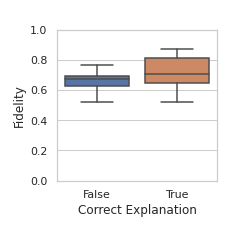}}
 \caption{no sing $\rightarrow$ sing \\ (waveform)}
\end{subfigure}%
\begin{subfigure}{0.5\columnwidth}
 \centerline{
 \includegraphics[width=\textwidth]{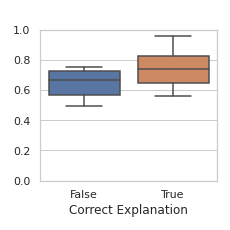}}
  \caption{no sing $\rightarrow$ sing \\ (spectrogram)}
\end{subfigure} 
\caption{%Comparison of f 
Fidelity scores for explanations that correctly and incorrectly identified the most relevant segment.}   
 \label{fig:fidelity}
\end{figure}

In the last experiment described above we have created a setting with $k=1$ where we exactly know which segment is relevant for the prediction, and we can use this information to compare whether the fidelity is different for ``correct'' and ``wrong'' explanations, i.e. explanations which match the added segments of the perturbation as opposed to explanations which do not.
As can be seen from Figure~\ref{fig:fidelity}, the fidelity for ``correct'' explanations is on average only slightly higher than for ``wrong'' predictions, with ranges of fidelity values strongly overlapping. Based on the fidelity score alone, a user hence cannot decide whether an explanation is ``correct'' and whether it shows the most relevant segment(s) for a prediction.

%----------------------------------------------------------------------

\section{Discussion and Conclusion}
\label{sec:conclusion}

In a nutshell, the central results of our investigations
can be summarised as follows:

\begin{itemize}\itemsep-0.2em
    \item Figure~\ref{fig:horses}: ``Obvious'' causes for predictions can be detected with high fidelity.
    \item Table~\ref{tbl:preview}: When computing the 3 most relevant interpretable features with LIME, we detect segments that are able to flip the label for only 21-53\% of the excerpts.
    \item Figure~\ref{fig:flipped}: When using a varying number of interpretable features, we detect fewer ``correct'' segments than when simply taking the same number of segments with the highest magnitude.
    \item Table~\ref{tbl:positive}: When using all interpretable features that received a positive weight, we detect fewer ``correct'' segments than when taking the same number of segments with the highest magnitude.
    \item Figure~\ref{fig:subset_perturbation}: In a setting where we only add partial perturbations, only few segments are correctly detected.
    \item Figure~\ref{fig:fidelity}: Based on the fidelity score it is impossible to judge the quality of an explanation.
\end{itemize}

Taken together, we believe that these results support the following conclusions: (1) local model-agnostic explanations cannot reliably detect the input regions most relevant for a prediction unless they are rather obvious;
(2) evaluation based on ``what looks reasonable'' leads to accepting explanations that do not reveal the real cause of a prediction;
(3) the fidelity score may give a false sense of security about the quality of explanations.

In \cite{goepfert2019recovering}, G\"opfert et al.\ write, ``It might be tempting to judge explanatory methods on whether they succeed in identifying features that a human observer thinks should be relevant to the classification, but the existence of adversarial examples shows that the reasoning of humans and neural networks can differ dramatically.'' This observation points to a fundamental dilemma between ``veridical'' explanations that faithfully reflect the workings of the classification model (and may not be accessible to model-agnostic explanation methods such as LIME), and ``human-interpretable'' explanations that would connect a machine decision to concepts familiar to us. Our results confirm that one should be careful in interpreting model-agnostic explanations as explanations of the underlying model, and that experiments with carefully crafted examples are important to get more detailed insights into the properties of such methods.

\section{Acknowledgements}
This work is supported by the Austrian National Science Fund (FWF, project P31988).

% For bibtex users:
\bibliography{ISMIR2021}

\end{document}